\documentstyle[12pt,psfig]{article}
\def\reference{\parskip 0pt\par\noindent\hangindent 0.5 truecm}

\newcommand{\ga}{\mathrel{\hbox{\rlap{\lower.55ex \hbox {$\sim$}}
 \kern-.3em \raise.4ex \hbox{$>$}}}}
\newcommand{\la}{\mathrel{\hbox{\rlap{\lower.55ex \hbox {$\sim$}}
 \kern-.3em \raise.4ex \hbox{$<$}}}}
\newcommand{\Lsun}{\mbox{${{\rm L}_\odot}$}}
\newcommand{\Msun}{\mbox{${{\rm M}_\odot}$}}
\newcommand{\Rsun}{\mbox{${{\rm R}_\odot}$}}

\begin{document}
%
%
\title{Helium-Star Mass Loss and its Implications for Black-Hole Formation
and Supernova Progenitors}
%


\author{Onno R.\ Pols$^{1,2}$ and
 Jasinta D.M.\ Dewi$^{3,4}$
} 

\date{}
\maketitle

{\center $^1$ Department of Mathematics, PO Box 28M, Monash
University, Vic 3800,
Australia\\[3mm]
$^2$ Astronomical Institute, Postbus 80000, 3508 TA Utrecht, The Netherlands
\\ o.r.pols@astro.uu.nl \\[3mm]
$^3$ Astronomical Institute, Kruislaan 403, 1098 SJ Amsterdam, The Netherlands
\\ jasinta@astro.uva.nl \\[3mm]
$^4$ Bosscha Observatory and Department of Astronomy, Lembang 40391,
Bandung, Indonesia \\[3mm]
}

%
\begin{abstract}
Recently the observationally derived stellar-wind mass-loss rates
for Wolf-Rayet stars, or massive naked helium stars, have been
revised downwards by a substantial amount. We present evolutionary
calculations of helium stars incorporating such revised mass-loss
rates, as well as mass transfer to a close compact binary
companion. Our models reach final masses well in excess of
10\,\Msun, consistent with the observed masses of black holes in
X-ray binaries. This resolves the discrepancy found with
previously assumed high mass-loss rates between the final masses
of stars which spend most of their helium-burning lifetime as
Wolf-Rayet stars ($\sim3$\,\Msun) and the minimum observed black
hole masses (6\,\Msun). Our calculations also suggest that there
are two distinct classes of progenitors for Type Ic supernovae:
one with very large initial masses ($\ga 35$\,\Msun), which are
still massive when they explode and leave black hole remnants, and
one with moderate initial masses ($\sim 12-20$\,\Msun) undergoing
binary interaction, which end up with small pre-explosion masses
and leave neutron star remnants.

\end{abstract}


\bigskip

%
%

\section{Introduction}
\label{s:intro}

A helium star is the naked core of a star that has lost its H-rich
envelope, as a result of either a strong stellar wind or binary
interaction.  In a very massive star (initial mass $M_{\rm i} \ga
40$\,\Msun) the stellar wind is strong enough to remove the
envelope before or during the central He-burning phase of
evolution. Such stars can thus leave single naked He-burning stars
with masses larger than about 15\,\Msun.  Less massive helium
stars can be produced by mass transfer in a close binary system,
if the primary (more massive) component of a binary has $M_{\rm i}
> 2-3$\,\Msun\ and the orbital dimensions are such that Roche lobe
overflow (RLOF) starts during the main sequence (case A mass
transfer) or during the Hertzsprung gap or the first giant branch
(case B). The remnant of mass transfer will then be an almost
naked helium star with $M > 0.32$\,\Msun\ (the minimum mass for
helium burning) orbiting a more massive main sequence star (van
den Heuvel 1994).  This allows quite a large range in orbital
periods as well as initial masses, and helium stars in binaries
are therefore expected to be quite common.\footnote{The fact that
not many such systems are known can be understood by considering
that the main sequence star completely dominates the spectrum at
optical wavelengths (Pols et al.\ 1991).}  Here we will consider
helium stars that are massive enough to undergo core collapse and
end their lives as neutron stars or black holes, i.e. $M_{\rm He}
\ga 2-2.5$\Msun. This requires initial masses of at least
$8-10$\,\Msun.

\subsection{Mass Loss and the Formation of Black Holes}

Helium stars with $M \ga 5-10$\,\Msun\ have strong mass loss
themselves; they are identified with hydrogen-free Wolf-Rayet (WR)
stars of type WN or WC. Mass loss strongly influences the
evolution of these stars, as well as their final masses and fate
(Langer 1989).  Unfortunately, the observationally derived
mass-loss rates for WR stars are very uncertain. Evolution models
for He stars computed with the mass-loss parametrisations
suggested by Langer (1989) or Hamann, Koesterke, \& Wessolowski
(1995) result in strong mass convergence: even for the largest
initial masses the final mass before core collapse is no more than
3 to 4\,\Msun\ (e.g.\ see Woosley, Langer, \& Weaver 1995;
Wellstein \& Langer 1999). However, recent WR wind models that
take into account the inhomogeneous structure of the wind have led
to a downward revision of the mass-loss rate by a factor of 3 to 5
(Hamann \& Koesterke 1998; Nugis \& Lamers 2000).  In fact, the
Nugis \& Lamers rate is nearly an order of magnitude smaller than
the Hamann et al.\ (1995) rate for WN stars.

Whether the helium star leaves a neutron star or a black hole
remnant will be determined to a large extent by its final mass
before core collapse, or rather, by its final core mass. However,
the outcome depends on the details of the explosion mechanism and
the limiting mass is very uncertain (e.g.\ see Fryer et al.\
2002).  Important observational constraints come from X-ray
binaries with low-mass companions (LMXB) in which the dynamically
inferred mass of the compact star exceeds 3\,\Msun, the maximum
possible neutron star mass.  A strict lower limit to the black
hole (BH) mass is set by the mass function, which in several LMXB
is at least 6\,\Msun\ (e.g.\ Casares, Charles, \& Naylor 1992;
McClintock et al.\ 2001).  In a few cases, the inferred BH mass is
very likely to be $\ga 10$\,\Msun\ (McClintock 1998; Orosz et al.\
2001).  In the evolutionary scenarios for the formation of
BH--LMXB the immediate progenitor of the black hole is a naked He
star. Because the black holes in these systems can hardly have
accreted any mass (King \& Kolb 1999), the pre-explosion mass must
have exceeded the BH mass, probably by a substantial amount if the
collapse was accompanied by a supernova explosion.  Clearly, if
all naked He stars reach final masses of only $3-4$\,\Msun\ these
facts cannot be accounted for.  One possible solution is that the
progenitors of the observed systems result from case C mass
transfer (i.e.\ RLOF started after central He exhaustion) rather
than case A/B (Brown, Lee, \& Bethe 1999). In that case the naked
He star has already gone through core He burning and is close to
core collapse when it forms, so there is insufficient time to lose
a significant amount of mass in a stellar wind.  Whether this
scenario can explain all (or any) of the observed BH-LMXBs depends
critically on the range of initial masses and orbital periods for
which case C mass transfer is possible, which in turn depends
quite sensitively on uncertain details of stellar evolution
models. However, in the light of the revised WR mass-loss rates,
it is worthwhile to reconsider whether case A/B mass transfer
cannot after all produce massive black holes.  Nelemans \& van den
Heuvel (2001), using a simple analytic estimate, suggest that this
is indeed possible. In this paper we present full-scale
evolutionary calculations incorporating the Nugis \& Lamers (2000)
mass-loss rate in order to investigate this question.

\subsection{Supernovae of Types Ib and Ic}

The amount of mass loss from a helium star, and hence its final
mass, also has important consequences for the type of supernova
(SN) explosion it produces.  Type Ib supernovae show helium lines
in their spectra, and their connection to the core collapse of
helium stars is quite straightforward. Type Ic supernovae show
little or no evidence for helium, and their progenitors are less
obvious.  However, the similarity of their late-time spectra
indicates that the progenitors of SN Ib and SN Ic are related, and
in fact small amounts of helium may be present in SNe Ic
(Filippenko, Barth, \& Matheson 1995).  Hence helium cores that
have been stripped of all or most of their helium layers are the
best candidates for SNe Ic.  As with the formation of helium stars
themselves, either strong stellar-wind mass loss or mass transfer
in a binary may be responsible for this additional stripping.  If
the initial mass is large enough, the strong WR wind can expose
the C-rich core and remove most of the helium.  In binary systems
that have gone through case A/B mass transfer, a second phase of
mass transfer is possible from the helium star to its companion
(case BB mass transfer), if the orbit is close enough and the
helium star is not too massive ($M \la 6$\,\Msun).

In this paper we attempt to identify likely progenitors of Type Ic
supernovae and of black holes in X-ray binaries in the light of
revised mass-loss rates for Wolf-Rayet stars.  To this end we
present evolutionary calculations of massive helium stars, losing
mass both through a stellar wind and by mass transfer to a nearby
compact binary companion. We describe the calculations and their
initial conditions in Section~\ref{s:calc}, and the relevant
results in Section~\ref{s:results}. The implications are discussed
in Section~\ref{s:concl}.

\section{Evolutionary Calculations and Initial Conditions}
\label{s:calc}

The stellar evolution calculations on which this paper is based
are described in detail by O.R.\ Pols (in preparation) and Dewi et
al.\ (2002). The initial configuration at the start of each
calculation is a homogeneous star of almost pure helium with a
solar fraction of heavier elements ($Y=0.98$, $Z=0.02$) which we
evolve through central helium burning and through carbon burning.
We consider initial helium star masses $M_{\rm He,i}$ between 2
and 32\,\Msun. The underlying assumption is that such a He star
has formed by case A or case B mass transfer in a binary, and we
consider two extreme cases.

In the first case (I) we assume that the companion is a main
sequence star and that mass transfer was conservative, which leads
to a widening of the orbit (van den Heuvel 1994).  For this case
we computed the evolution of the He star in isolation, i.e.\ not
considering the possibility of a second (case BB) mass transfer
phase as the He star expands. This will be correct except perhaps
for He stars less massive than 2.5\,\Msun\ which can attain radii
of 100\,\Rsun\ or more.  In the calculations we apply the
stellar-wind mass-loss rate of Nugis \& Lamers (2000), which
depends on the luminosity $L$ and on the surface abundances, thus:
\begin{equation}
\dot{M} = 1.0\times10^{-11} (L/\Lsun)^{1.29}\, Y^{1.7}\, Z^{0.5} \quad
{\rm \Msun/yr} .
\end{equation}
The composition dependence causes mass loss to accelerate when C
and O are exposed at the stellar surface, which has interesting
consequences.  This set of single-star models applies both to
sufficiently wide case A/B binaries, and to actually single He
stars with $M_{\rm He,i} \ga 15$\,\Msun, which can be formed by
stellar-wind mass loss.  These calculations are described by O.R.\
Pols (in preparation).

In the second case (II) we assume that the companion is a neutron
star (NS) in a close orbit, and we compute the non-conservative
mass transfer that ensues when the He star fills its Roche lobe.
We assume that the NS accretes up to its Eddington limit and that
the excess mass is lost from the system with the specific orbital
angular momentum of the NS. Such a system can form as a result of
the spiral-in of the neutron star in the envelope of a massive
star evolving off the main sequence in an initially wide orbit,
i.e.\ as a remnant of a Be/X-ray binary (van den Heuvel 1994). We
consider a range of orbital periods for the He-star + NS binary,
consistent with the (very uncertain) periods expected after such a
spiral-in (Dewi et al.\ 2002). This case can be considered as an
advanced evolutionary stage of case I, i.e.\ after the
first-formed He star has undergone core collapse and become a
neutron star.  In these calculations we apply a somewhat different
mass-loss rate, namely one fourth of the $L$--dependent rate
proposed by Hamann et al.\ (1995).  This rate is larger (by
roughly a factor of two) than the Nugis \& Lamers rate, but there
is no real discrepancy with case I because the final mass in case
II is determined by RLOF and not by the stellar wind for the
masses considered (up to 6.6\,\Msun; more massive He stars expand
very little and usually avoid RLOF).  We take full account of the
orbital evolution resulting from non-conservative RLOF, the
stellar wind, and gravitational wave radiation (for details see
Dewi et al.\ 2002).

We note that other cases are possible, e.g.\ non-conservative case
B mass transfer leading to a He-star + main sequence binary in a
fairly close orbit.  In such a system we also expect the He star
to undergo case BB mass transfer, but this time to a main sequence
star and
--- in all likelihood --- conservatively. The result would probably
be intermediate between cases I and II, but this should be borne
out by actual calculations.

\section{Results: Final Masses and Helium Amounts}
\label{s:results}

In this section we concentrate on the final configurations
resulting from the calculations: the stellar mass $M_{\rm He,f}$
and the amount of helium left in the envelope $\Delta M_{\rm
He,f}$ just before core collapse,\footnote{The values given refer
to the end of the calculations, which in most cases extended well
into carbon shell burning.  During the very short remaining time
until core collapse, the masses cannot change significantly and
can indeed be considered as final.} and we discuss their
implications for supernovae and black hole formation.  For other
aspects of the models we refer to the papers by O.R.\ Pols (in
preparation) and Dewi et al.\ (2002).

\subsection{Case I}

\begin{figure}[b]
\centerline{\psfig{file=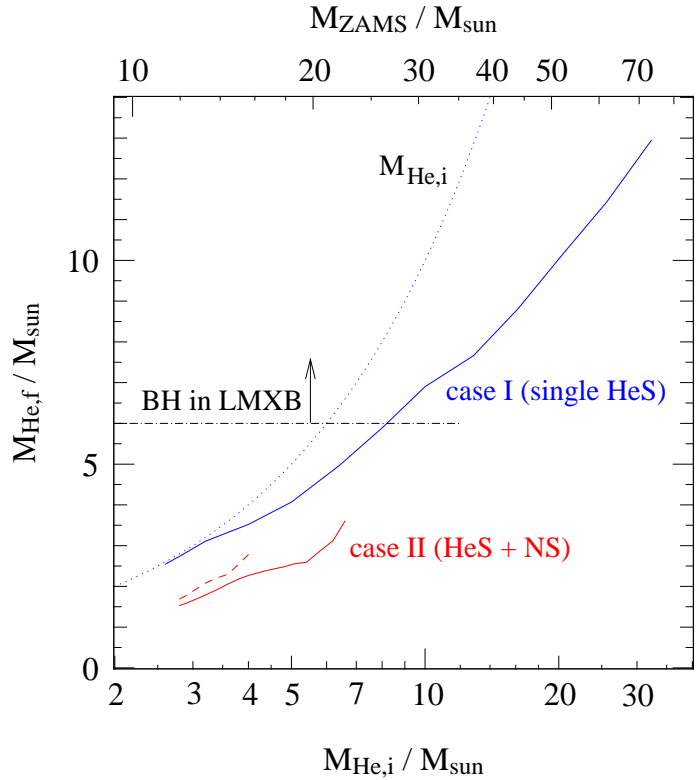,height=11cm}} \caption{The
relation between initial helium star mass and final mass, for wide
binaries (effectively single He stars, case I) and close binaries
with additional case BB mass transfer to a NS companion (case II).
Stellar-wind mass loss according to the prescription by Nugis \&
Lamers (2000) has been applied for case I (upper, blue solid
line).  The lower, red solid line is for case II with binary
periods of 0.08 to 0.09 days, the dashed red line is for periods
of 0.4 to 0.5 days.  The observed lower mass limit for the most
massive black holes in LMXBs is shown as a dashed--dotted line.
The dotted line reproduces the initial He-star mass (note that the
horizontal scale is logarithmic). Along the top the ZAMS mass has
been indicated, assuming that the He star formed as a result of
case B mass transfer.} \label{f:finalmass}
\end{figure}

\begin{figure}[b]
\centerline{\psfig{file=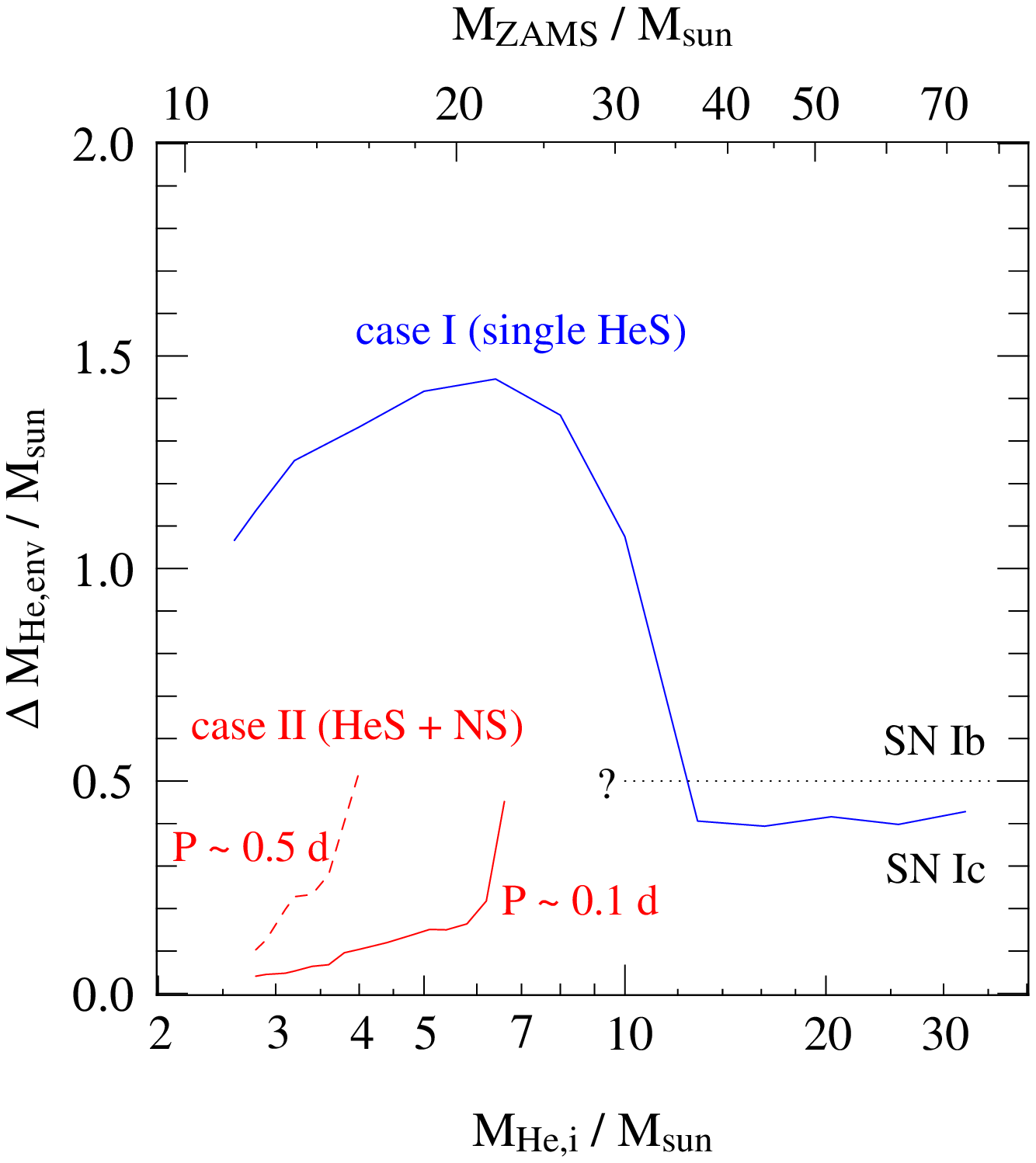,height=11cm}} \caption{Final
mass of helium left in the envelope of the He star before core
collapse. The curve styles are the same as in
Figure~\ref{f:finalmass}. The dotted line suggests a possible
critical mass of He, below which the explosion would appear as a
Type Ic supernova.} \label{f:envmass}
\end{figure}

The final masses of the single He-star models (i.e.\ remnants of
conservative case A/B mass transfer) are plotted in
Figure~\ref{f:finalmass}. At the top of the figure, the initial
zero-age main sequence (ZAMS) mass has been indicated, under the
assumption that the initial mass of the He star equals the core
mass at helium ignition.  This is approximately correct for case
B, but underestimates the ZAMS mass if the He star formed by case
A.  The relation $M_{\rm He,i} = 0.098 {M_{\rm ZAMS}}^{1.35}$ has
been used to estimate the ZAMS mass (Hurley, Pols, \& Tout 2000,
equation~44). Final masses up to 13\,\Msun\ are reached for the
most massive stars considered. Although the final--initial mass
relation levels off, no mass convergence is apparent.  We see that
for case B binaries starting with $M_{\rm ZAMS} > 27$\,\Msun,
final masses in excess of 6\,\Msun\ are reached.  This is
consistent with the largest observed black hole masses if no
additional mass is lost during the collapse of the black hole.  On
the other hand if we assume, for example, that 25 per cent of the
mass is ejected when the black hole forms, the ZAMS mass needs to
be greater than 40\,\Msun.  Under this assumption it is still
possible to obtain even a 10\,\Msun\ black hole after case B mass
transfer, starting from a 32\,\Msun\ He star (or $M_{\rm ZAMS}
\approx 70$\,\Msun).

A slight break in the initial--final mass relation is apparent
between $M_{\rm He} = 10$ and 12\,\Msun. This is caused by the
composition dependence in the Nugis \& Lamers mass-loss rate which
increases with $Z$, the fraction of heavy elements.  For initial
mass up to 10\,\Msun, the products of He burning are never exposed
to the surface, while for larger masses the surface becomes carbon
enriched (these stars make the transition from WN to WC stars). As
a result the mass loss accelerates when this transition is made.

In Figure~\ref{f:envmass} we show the final mass of $^4$He left in
the envelope prior to core collapse, as a function of initial He
star mass. For case I two regimes can be distinguished.  For
$M_{\rm He,i} \la 10$\,\Msun\ more than 1\,\Msun\ of He remains in
the envelope, while for $M_{\rm He,i} \ga 12$\,\Msun\ only
$\approx0.4$\,\Msun\ of He is left. The transition is quite sharp,
for the same reason as indicated above: mass loss accelerates when
the star becomes a WC star and removes a large fraction of the
remaining helium.  The transition is smoother if mass loss does
not depend on composition but only on luminosity, as in the
calculations of Wellstein \& Langer (1999, their Figure~6).
Following the arguments in that paper, it is tempting to identify
progenitors of Type Ib supernovae with $M_{\rm He,i} \la
10$\,\Msun\ and those of Type Ic SNe with $M_{\rm He,i} \ga
12$\,\Msun.  Although the latter still have a substantial amount
of helium, this may not show up in the SN spectrum if it is not
mixed with the radioactive $^{56}$Ni (Woosley \& Eastman 1997).
Such Type Ic SN progenitors would be massive ($> 7$\,\Msun), and
probably leave black hole remnants, although in view of the large
uncertainties in the core-collapse mechanism, neutron star
remnants cannot be excluded. Although rather massive black holes
may form by direct collapse (Fryer 1999), in at least one LMXB
there is strong evidence that the formation of the black hole was
indeed accompanied by a supernova explosion (Israelian et al.\
1999). These explosions can possibly be identified with hypernovae
such as SN 1998bw, which was of Type Ic and associated with a
$\gamma$-ray burst.  A massive, almost bare CO core in combination
with a large explosion energy can explain the bright and slowly
declining lightcurve and broad spectral features of such
hypernovae (Iwamoto et al.\ 2000).

\subsection{Case II}

Helium stars in close orbits around a NS companion initially also
lose mass in a stellar wind.  When non-conservative RLOF starts,
much higher mass-loss rates are reached, up to
$10^{-4}$\,\Msun/yr, because mass transfer occurs on the thermal
timescale of the He star (see Dewi et al.\ 2002 for more details).
At the end of the calculations most of the envelope has been
transferred and lost from the binary system, unless RLOF starts
when the He star is already close to carbon burning.  As a
consequence, the final masses, shown in Figure~\ref{f:finalmass},
depend somewhat on the initial orbital period of the He-star + NS
system, but are usually close to the CO core mass except for the
widest orbits and most massive He stars.  In all cases they are
substantially smaller (1.5 -- 3\,\Msun) than in Case I, where only
the stellar wind operates.  For $M_{\rm He,i} > 6.6$\,\Msun\ RLOF
becomes dynamically unstable, but such systems are rare because He
stars of such masses hardly expand.  The small final mass implies
core collapse will result in neutron star formation. Unless the
explosion disrupts the binary, the remnants of these systems are
therefore double NS binaries which, if close enough to merge in a
Hubble time, are candidate progenitors of $\gamma$-ray bursts.

The final mass of He in the envelope, as shown in
Figure~\ref{f:envmass}, is also much smaller than in case I but,
like the final stellar mass, depends on the initial He star mass
and on the orbital period.  In most cases it is $\la 0.2$\,\Msun,
and for the shortest orbital periods, $P \approx 0.1$\,day, it can
be as small as 0.04\,\Msun.  This is much less than is achieved
for massive He stars by stellar wind only (case I, see above).  It
is also less than in the conservative case BB mass-transfer models
by Wellstein \& Langer (1999).  We conclude that non-conservative
case BB mass transfer to a compact companion is the most efficient
way to produce almost bare CO cores prior to explosion.  These
stars will almost certainly produce a Type Ic supernova. The small
progenitor mass would result in a relatively faint, fast declining
lightcurve.  Such a model was first suggested by Nomoto et al.\
(1994) as the progenitor of the Type Ic SN 1994I, and shown to
match the observed lightcurve (Iwamoto et al.\ 1994).

\section{Summary and Conclusions}
\label{s:concl}

If the reduced and composition-dependent mass-loss rate for WR
stars of Nugis \& Lamers (2000) is adopted, the following picture
emerges.  Binary components with $M \ga 35$\,\Msun\ that form WR
stars through case A or B mass transfer, as well as single stars
massive enough to form WR stars directly, reach final He-star
masses in excess of 7\,\Msun\ and have only a small amount of He
($\sim0.4$\,\Msun) left in their envelopes.  Such stars probably
leave black hole remnants, and the final masses are large enough
to be consistent with the observed BH masses in X-ray binaries.
Unless such black holes form by direct collapse, these stars very
likely undergo a Type Ic supernova explosion which can possibly be
identified with bright, slowly declining hypernovae such as SN
1998bw.

Less massive stars in binaries undergoing case A/B mass transfer, but wide
enough to avoid case BB mass transfer, have smaller final masses and at
least 1\,\Msun\ of He left in their envelope.  These stars undergo a Type
Ib supernova and leave either neutron stars or, possibly, black holes
for stars at the upper end of the mass range.

Binary components with initial mass $\la 20$\,\Msun\ and in close
enough orbits undergo case BB mass transfer.  Independently of the
adopted WR mass-loss rate, this results in even smaller final
masses and a smaller amount of He left in the envelope. The most
efficient way to end up with an almost He-free star prior to core
collapse is by case BB mass transfer to a neutron star in a very
close orbit (i.e.\ RLOF from the initial secondary component of
the binary, after the primary has already collapsed).  For small
initial masses and orbits with $P\sim0.1$\,day it is possible to
have as little as 0.04\,\Msun\ of He left in the envelope.  Such
stars almost certainly explode as a SN Ic, and have very small
pre-SN masses (1.5 to 3\,\Msun).  They can be identified with
faint, fast declining Type Ic supernovae like SN 1994I.

In summary, our most important conclusions are:
\begin{itemize}
\item Adoption of the Nugis \& Lamers mass-loss rate leads to final He-star
masses after case A/B mass transfer that are consistent with the observed
black hole masses. There is no need to resort to case C mass transfer
to explain the observed BH binaries, although such a scenario remains a
possibility.
\item Massive binary evolution leads to two distinct classes of SN Ic
progenitors: one with large pre-explosion mass, $M>7$\,\Msun,
formed from initial masses $M>35$\,\Msun, which probably leave BH
remnants and can possibly be identified with hypernovae, and the
other with small pre-explosion mass, 1.5 -- 3\,\Msun, formed by
case BB mass transfer in binaries with $M\sim12 - 20$\,\Msun,
which leave NS remnants.
\end{itemize}
The latter conclusion is similar to that of Wellstein \& Langer (1999), but
in their models massive SN Ic progenitors have much smaller pre-explosion
masses, $3-4$\,\Msun.

We note that other consequences of the mass-loss rate of WR stars also need
to be explored, in particular for the properties of WR stars
themselves. Their luminosity and abundance distributions and the number
ratio of WN to WC stars all depend on the adopted mass-loss rate. This is
beyond the scope of this paper, but we emphasise that in order to draw
definite conclusions about WR mass loss and BH formation, all these aspects
have to be considered in conjunction.

\section*{Acknowledgments}

It is a pleasure to thank Norbert Langer, Ed van den Heuvel, Gijs
Nelemans, and Gert-Jan Savonije for stimulating discussions and
valuable suggestions. Jasinta Dewi acknowledges sponsorship by NWO
Spinoza Grant 08-0 to E.P.J.\ van den Heuvel.

\section*{References}






\reference Brown, G.E., Lee, C.H., \& Bethe, H.A.\ 1999, NewA, 4,
313

\reference Casares, J., Charles, P.A., \& Naylor, T.\ 1992,
Nature, 355,
  614

\reference Dewi, J.D.M., Pols, O.R., Savonije, G.J., \& van den
Heuvel,
  E.P.J.\ 2002, MNRAS, in press (astro-ph/0201239)

\reference Filippenko, A.V., Barth, A.J., \& Matheson, T.\ 1995,
ApJ,
  450, L11

\reference Fryer, C.L.\ 1999, ApJ, 522, 413

\reference Fryer, C.L., Heger, A., Langer, N., \& Wellstein, S.\
2002, ApJ, submitted (astro-ph/0112539)

\reference Hamann, W.R., \& Koesterke, L.\ 1998, A\&A, 335, 1003

\reference Hamann, W.R., Koesterke, L., \& Wessolowski, U.\ 1995,
A\&A, 299, 151

\reference Hurley, J.R., Pols, O.R., \& Tout, C.A.\ 2000, MNRAS,
315, 543

\reference Israelian, G., Rebolo, R., Basri, G., Casares, J.,
Martin, E.L.\ 1999, Nature, 401, 142

\reference Iwamoto, K., Nomoto, K., Hoflich, P., Yamaoka, H.,
Kumagai, S., \& Shigeyama, T.\ 1994, ApJ, 437, L115

\reference Iwamoto, K., et al.\ 2000, ApJ, 534, 660

\reference King, A.R., \& Kolb, U.\ 1999, MNRAS, 305, 654

\reference Langer, N.\ 1989, A\&A, 220, 135

\reference McClintock, J.E.\ 1998, in Accretion Processes in
Astrophysical
  Systems, AIP Conf.\ Proc.\ 431, ed.\ S.S.\ Holt \& T.R.\ Kallman (New
  York: AIP), 290

\reference McClintock, J.E., et al.\ 2001, ApJ, 555, 477

\reference Nelemans, G., \& van den Heuvel, E.P.J.\ 2001, A\&A,
376, 950

\reference Nomoto, K., Yamaoka, H., Pols, O.R., van den Heuvel,
E.P.J.,
  Iwamoto, K., Kumagai, S., \& Shigeyama, T.\ 1994, Nature, 371, 227

\reference Nugis, T., \& Lamers, H.J.G.L.M.\ 2000, A\&A, 360, 227

\reference Orosz, J.A., et al.\ 2001, ApJ, 555, 489

\reference Pols, O.R., Cot\'e, J., Waters, L.B.F.M., \& Heise, J.\
1991, A\&A, 241, 419


\reference van den Heuvel, E.P.J.\ 1994, in Interacting Binaries,
  Saas-Fee Advanced Course 22, ed.\ H.\ Nussbaumer \& A.\ Orr (Berlin:
  Springer-Verlag), 263

\reference Wellstein, S., \& Langer, N.\ 1999, A\&A, 350, 148

\reference Woosley, S.E., \& Eastman, R.G.\ 1997, in Thermonuclear
  Supernovae, NATO ASI Series C, Vol.\ 486, ed.\ P.\ Ruiz-Lapuente,
  R.\ Canal, \& J.\ Isern (Dordrecht: Kluwer), 821

\reference Woosley, S.E., Langer, N., \& Weaver, T.A.\ 1995, ApJ,
448, 315


\end{document}